\documentclass[a4paper,twocolumn,11pt,preprint]{quantumarticle}
\pdfoutput=1
\usepackage[utf8]{inputenc}
\usepackage[english]{babel}
\usepackage[T1]{fontenc}
\usepackage{amsmath}
\usepackage{hyperref}
\usepackage[numbers]{natbib}
\usepackage{physics}
\usepackage{amsmath}
\usepackage{amsfonts}
\usepackage{amssymb}
\usepackage{bbm}
\usepackage{color}
\usepackage{float}
\usepackage{booktabs}

\begin{document}

\title{Bosonic fields in states with  undefined particle numbers possess detectable non-contextuality features, plus more}

\author{Konrad Schlichtholz}
\email{konrad.schlichtholz@phdstud.ug.edu.pl}
\orcid{0000-0001-8094-7373}
\affiliation{International Centre for Theory of Quantum Technologies (ICTQT),
University of Gdansk, 80-308 Gdansk, Poland}
\author{Antonio Mandarino}
\email{antonio.mandarino@ug.edu.pl}
\orcid{0000-0003-3745-5204}
\affiliation{International Centre for Theory of Quantum Technologies (ICTQT),
University of Gdansk, 80-308 Gdansk, Poland}
\author{Marek \.Zukowski}
\orcid{0000-0003-3745-5204}
\affiliation{International Centre for Theory of Quantum Technologies (ICTQT),
University of Gdansk, 80-308 Gdansk, Poland}

\begin{abstract}
Most of the paradoxical, for the classical intuition, features of quantum theory were formulated for situations which involve a fixed number of particles. While one can now find a formulation of Bell's theorem for quantum fields, a Kochen-Specker-type reasoning  is  usually formulated for just one particle, or like in the case of Peres-Mermin square for two. A question emerges. Is it possible to formulate a contextuality proof for situation in which the numbers of particles are fundamentally undefined? We address this problem for bosonic fields.
We introduce a  representation of the $\mathfrak{su}(2)$ algebra in terms of boson number states in two modes that allows us to assess  nonclassicality of states of bosonic fields. As a figure of merit of  a nonclassical behaviour we analyze first of all contextuality, and we show that the introduced observables are handy and efficient to reveal violation of local realism, and to formulate entanglement indicators. We construct a method which extends the Kochen-Specker contextuality to bosonic quantum fields. A form of an inequality is derived using  a suitable  version of the Peres-Mermin square. The entanglement indicators use a witness built with specially defined Pauli-like observables. Finally, Bell-nonclassicality is discussed:  an inequality that involves the expectation  values of pairs of the Pauli-like operators is presented. The introduced  indicators are shown to be effective, e.g. they reveal nonclassicality in situaations involving undefined boson numbers. This is shown via quantum optical examples of the $2\times 2$ bright squeezed vacuum state,  and a recently discussed bright-GHZ state resulting from multiple three photon emissions in a parametric process.
\end{abstract}

\maketitle

\section{Introduction}
Quantum phenomena without any classical analogue are the fuel of many 
protocols of interest in quantum technologies, constituting the main resources for their different applications. To name a few, contextuality has been addressed as the key ingredient providing the speed-up in quantum computation over the classical one \cite{ContextualityAdv}, the violation of a Bell inequality enables one to implement  secure cryptographic schemes \cite{DeviceindepQKD} and entangled probes allow one to achieve better precision in metrological tasks \cite{Giovannetti_metr}.

In parallel, research on the quantum properties of bosonic fields, notably first of all quantum light, has constituted one of the pillars in the development of quantum technologies, and it is still of paramount importance.
Nonclassicality of a state of a quantum field can be 
revealed in many ways, ranging from looking at the negativity or singularity of a certain quasi-distribution function \cite{mandel_wolf_1995}, to criteria to establish the nonseparability of a multimode field \cite{Pan2012}. 
Following the approaches borrowed from finite-dimensional systems, another strategy to show whether a field admits a classical 
description or not is to introduce a set of inequalities 
that are fulfilled by measurement probabilities which are possible to model by a classical local-realistic theory, 
but are violated when taking into account the quantum mechanical predictions. 

If the local hidden variable theory imposes some constraints
on the structure of the joint probabilities or on the statistical correlations of measurements on multipartite systems, then these constraints can usually be put in a form of linear Bell inequalities. 
While if we consider hidden variable theories in which predetermined measurement values of any (degenerate) observable do not depend on the context in which the observable is measured, then the constraints are of  Kochen-Specker type. A context is defined by a non-degenerate observable, which commutes with the observable in question, see e.g. \cite{Mermin1993}. 

Here, we focus on the quantification of different forms of nonclassicality for optical fields introducing a suitable set of operators defined in terms of the number states of two modes of the field. Their mathematical expressions are handy and allow us to map some results derived for qubit-systems to optical states. We remark here that, despite our efforts, we still do not have a clear picture of how such measurements can be implemented in an experimental scenario.  This poses an interesting experimental challenge. 


\section{Pauli-like operators for bosonic fields}
\label{sec:operators}
Let us  introduce a set of Hermitian operators for optical fields which generalize the Pauli matrices in a Fock space describing two distinguishable modes of a bosonic field:

\begin{align}
\begin{split}
&\hat G_0=\mathbbm{1}-\sum_n\ket{n,n}\bra{n,n},
\end{split}\\
\begin{split}
&\hat G_1= \sum_{n\neq m}\ket{n,m}\bra{m,n},
\end{split}\\
\begin{split}
&\hat G_2=-i \, \textrm{sign} (\hat a^\dagger \hat a -\hat b^\dagger \hat b)\sum_{n\neq m}\ket{n,m}\bra{m,n},
\end{split}\\
\begin{split}
&\hat G_3=\textrm{sign} (\hat a^\dagger \hat a -\hat b^\dagger \hat b),
\end{split}
\end{align}
where $\hat a^\dagger, \hat b^\dagger$ are creation operators of two orthogonal bosonic modes, $\mathbbm{1}$ is the two mode identity, 
and $\ket{n, m}=\frac{1}{\sqrt{n!m!}}a^{\dagger n} b^{\dagger m} \ket{\Omega}$, with $\ket{\Omega}$ being the vacuum. The operator $\hat G_3$ coincides with one of sign Stokes operators introduced in \cite{schlichtholz2021simplified}. The operators $\hat G_i$ can also be written in more compact form:
\begin{equation}
\hat G_i= \left( \begin{array}{cc}\hat S_R^\dagger & \hat P_R^\dagger\end{array} \right)\sigma_i\left( \begin{array}{c}\hat S_R \\ \hat P_R\end{array} \right)=\overrightarrow{V}^\dagger\sigma_i\overrightarrow{V},\label{innaforma}
\end{equation}
where $\sigma_i$ is the $i$-th Pauli matrix, $\overrightarrow{V}^\dagger=\left( \begin{array}{cc}\hat S_R^\dagger & \hat P_R^\dagger\end{array} \right) $ and
\begin{align}
&\hat P_R=\sum_{m>n}\ket{n,m}\bra{n,m}, \\
&\hat S_R=\sum_{m>n}\ket{n,m}\bra{m,n}.
\end{align}
These operators fulfill the same commutation relation as Pauli matrices:
\begin{equation}
[\hat G_i,\hat G_j]=2i\epsilon_{ijk}\hat G_k,\label{commR}
\end{equation}
where $i,j,k=1,2,3$. They also fulfill the anticommutation relation: 
\begin{equation}
\lbrace\hat G_i,\hat G_j\rbrace=2\delta_{ij}\hat G_0,\label{anti}
\end{equation}
and from that follows (for details, see Appendix \ref{comm}): 
\begin{equation}
\hat G_i\hat G_j=\delta_{ij}\hat G_0+i\epsilon_{ijk}\hat G_k.\label{prodG}
\end{equation}
The $\hat G_0$ operator plays an analogous role of the identity, i.e. $\hat \sigma_0=\mathbbm{1},$ and obviously it commutes with all the other $G_i$ operators. Since these operators satisfy the commutation relation in (\ref{commR}), 
they are a representation of the two-boson $\mathfrak{su}(2)$ algebra. 
Such a representation has advantages over the standard two mode representation of $\mathfrak{su}(2)$ \cite{algebra_su2} given by:
\begin{equation}
\hat S_i=\frac{1}{2}A^\dagger\sigma_i A, \label{stokes}   
\end{equation}
where $A^\dagger=(\hat a^\dagger,\hat b^\dagger)$. It fulfills the anticommutation relation (\ref{anti}), in contrast to the Stokes operators (\ref{stokes}). This propriety turns to be crucial in a generalization of the Peres-Mermin square. Moreover, the $\hat G_i$ operators  are bounded (see Appendix \ref{spectrum}) contrary to those previously introduced in (\ref{stokes}). We remark that the new operators {\it under unitary transformation of the modes} do not form an equivalent $\mathfrak{su}(2)$ representation. In fact, given two sets of new operators, $G_i$ and $G'_i$, for different modes basis namely $\hat a,\hat b$ and $\hat a',\hat b'$, they are not linked through unitary transformation of the modes (see Appendix \ref{ap:rot}) as it is for the Stokes operator defined in (\ref{stokes}).

\section{Contextuality for quantum optical fields}

The Kochen-Specker theorem \cite{KS_th} asserts that quantum mechanics is not compatible with any local hidden variable noncontextual theory. Here, noncontextuality means that the value for an observable predicted by such a theory is independent of the experimental context, namely the set of  other co-measurable observables are simultaneously measured. In classical physics, the measurement outcomes of several observables do not depend on the order in which  such quantities are measured. At the same time, quantum mechanics forbids existence of joint probability distributions for non-compatible observables. Noncontextual description of quantum measurements has been experimentally falsified for  systems with a definite number of particles \cite{Context_exp1, Context_exp2}. However, few attempts have been proposed till now to quantify and measure it for systems allowing states with an undefined number of particles, such as quantum optical fields \cite{Asadian:Contesxtuality, Contex_CV}. 
Finding such examples is not an academic question, as one can imagine that in future there might be quantum informational processes based
on such systems and such states.

\subsection{Peres-Mermin square for optical observables}

We introduce a version of the Peres-Mermin square \cite{PERES:PM, Mermin:PM} for the $2\times2$ mode states of quantum optical fields, given that the operators $\hat G_i$ fulfill (\ref{prodG}) as Pauli matrices do. Following the convention already introduced in \cite{Asadian:Contesxtuality}, we denote by $\hat A_{pq}$ observables such that if two of those observables share a common index, they commute, i.e., the order in which they are measured does not have an impact on the outcome. 
We chose nine of such observables by mapping the Peres-Mermin square in the following way $\hat \sigma_i^1\otimes\hat\sigma_j^2\rightarrow \hat G_i^1\hat G_j^2$, where superscripts denote the measuring parties. The Peres-Mermin square for the set of observables of the quantum optical fields reads:

\begin{table}[H]
\begin{center}
\begin{tabular}{@{}llll@{}}
\toprule
$\hat A_{ij}$ & $i=1$                  & $i=2$                  & $i=3$                  \\ \midrule
$j=1$    & $\hat G_3^1\hat G_0^2$ & $\hat G_0^1\hat G_3^2$ & $\hat G_3^1\hat G_3^2$ \\
$j=2$    & $\hat G_0^1\hat G_1^2$ & $\hat G_1^1\hat G_0^2$ & $\hat G_1^1\hat G_1^2$ \\
$j=3$    & $\hat G_3^1\hat G_1^2$ & $\hat G_1^1\hat G_3^2$ & $\hat G_2^1\hat G_2^2$ \\ \bottomrule
\end{tabular}
\end{center}
\end{table}
Following the method proposed in \cite{Cabello:Contextuality} to test whether quantum mechanics violates the bound imposed by a noncontextual hidden variable theory, we consider the following operator:
\begin{align}
\begin{split}
&\hat O=\hat A_{11}\hat A_{12}\hat A_{13}+\hat A_{21}\hat A_{22}\hat A_{23}+\hat A_{31}\hat A_{32}\hat A_{33}\\
+&\hat A_{11}\hat A_{21}\hat A_{31}+\hat A_{12}\hat A_{22}\hat A_{32}-\hat A_{13}\hat A_{23}\hat A_{33},\label{inOp}
\end{split}
\end{align}
where, in each term, one index is shared among the three operators $\hat A_{pq}$.
We make here an observation: the observables we have introduced are not dichotomic, whereas they have spectrum $\{-1, 0, 1\}$. This is in contrast to the observable used in the original construction of the Peres-Mermin square, namely the spin$-\frac12$ observables. However, the maximal average value of the expression (\ref{inOp}) in a noncontextual hidden variable (NCHV) model computed via the method given in \cite{Cabello:Contextuality} yields the same value $4$, both for dichotomic observables and for our generalized Pauli-observable for two radiation modes. Thus, we obtain the inequality:

\begin{align}
\begin{split}
&\langle A_{11} A_{12} A_{13}+ A_{21} A_{22} A_{23}+ A_{31} A_{32} A_{33}\\
+& A_{11} A_{21} A_{31}+ A_{12} A_{22} A_{32}- A_{13} A_{23} A_{33}\rangle_{NCHV}\leq 4.\label{inequality}
\end{split}
\end{align}
For the quantum case we observe that the first five terms of (\ref{inOp}) have the same form $\hat G_0^1\hat G_0^2$ and the last term is equal to $-\hat G_0^1\hat G_0^2$. Thus, as expected $\mathbbm{1}\rightarrow\hat G_0$. The operator $\hat G_0^1\hat G_0^2$ is a projection into the subspace of states having a different number of photons in both modes per each part. 
The subspace of such states is spanned by vectors $\ket{k,l;m,n}$ which are $a^{\dagger k}_1b^{\dagger l}_1a^{\dagger m}_2b^{\dagger n}_2\ket{\Omega}$ after normalization. The lower indices of creation operators denote measurement parties, $\ket{\Omega}$ is the vacuum state and $k\neq l$, $n\neq m$. Meanwhile, states having the same number of particles in both modes in one of the parties i.e. $k=l\vee n=m$ are in its orthogonal complement, to which we will refer as the \textit{diagonal subspace}.  As a consequence, the expectation value of such an operator for a generic state $\ket{\psi}$ is 
\begin{equation}
\label{eq:mean_G0}
    \langle\hat G_0^1\hat G_0^2\rangle_\psi =1-P(d|\psi),
\end{equation} 
where $P(d|\psi)$ is the probability that we get as outcome an eigenvalue corresponding to a state belonging to the diagonal subspace when measuring $\hat G_0^1\hat G_0^2$ on the state $\ket{\psi}$. We can write the expectation value of the operator defined in Eq.\ref{inOp} as a function of the probability appearing in Eq. \ref{eq:mean_G0}: 
\begin{equation}
\langle\hat O\rangle_\psi=6-6P(d|\psi).
\end{equation}
Thus, we can detect contextuality if:
\begin{equation}
P(d|\psi)<\frac{1}{3},\label{prob13}
\end{equation}
as we have a violation of the inequality (\ref{inequality}).
A pitfall of this approach is that we lose state independence when considering the inequality violation. Again, the reason relies on the spectrum of our observables and, in particular, on the existence of the zero eigenvalue. However, we can still test the contextuality of a large class of states useful for the implementation of quantum optical technologies. 

We stress that such construction to generalize the Peres-Mermin square using  standard Stokes operators ($2 \hat S_i$) and their normalized modification \cite{ZUKUPRA} fails to give satisfactory results due to the fact that both version of Stokes operators do not fulfill the anticommutaion relations of Pauli matrices. 
\subsection{Bright Squeezed Vacuum case}
To substantiate our approach, we check when we observe a violation of the  inequality in (\ref{inequality})  considering as paradigmatic example the $2\times 2$ Bright Squeezed Vacuum (BSV). This state consists of two orthogonal modes per part and it can be written in the form:
\begin{align}
\begin{split}
\ket{\psi_-}&=
\frac{1}{\cosh^2(\Gamma)}\sum_{n=0}^\infty \frac{\tanh^n(\Gamma)}{n!} (a_1^\dagger b_2^{\dagger}-a_2^\dagger b_1^{\dagger})^n \ket{\Omega}
\\&=
\frac{1}{\cosh^2(\Gamma)}\sum_{n=0}^\infty\sqrt{n+1}\tanh^n(\Gamma)\ket{\psi^n},\label{eq:bsv}
\end{split}
\end{align}
where $a^\dagger_X, b^\dagger_X (a_X, b_X)$ are the creation (annihilation) operators of the two orthogonal modes assigned to one of the two beams transmitted to the measurement part $X=1,2$. Furthermore, the parameter $\Gamma$ denotes the amplification gain and it is related to the power of the coherent source impinging on the nonlinear crystal in the parametric down conversion process and 
\begin{multline}
\label{eq:psi_n}
\ket{\psi_-^n}=\frac{1}{\sqrt{n+1}}\sum_{m=0}^n\\(-1)^m\ket{(n-m)_{a_1},m_{b_1};m_{a_2},(n-m)_{b_2}}.   
\end{multline}
We note that when $n$ is even, in the superposition in (\ref{eq:psi_n}) there is always one term that belongs to the diagonal subspace, namely the one having $(n-m)_{a_x}=m_{b_x}$, where $x=1,2$. In contrast, for $n$ odd, there are never terms from the diagonal subspace. 
Thus, by means of (\ref{eq:bsv}) the conditional probability of observing a result that can be associated with a state belonging to the diagonal subspace whenever a bright squeezed vacuum state is prepared reads:
\begin{equation}
P(d|BSV)=\frac{1}{\cosh^4(\Gamma)}\sum_{n=0}^\infty\tanh^{4n}(\Gamma)=\sech(2\Gamma).
\end{equation}  
From that follows that for $\Gamma>0.89$ the BSV state does not allow a noncontextual hidden variable description, because in this range inequality (\ref{prob13}) is fulfilled. In the limit of high pump power, namely for $\Gamma\rightarrow \infty,$ we have $P(d|BSV)=0$. Thus, in the limit of mean photon number tending to infinity we obtain the same violation as in the case of qubits, which we shall address in the next section.

\subsection{Qubit case}
The original formulation of the Peres-Mermin square was in terms of qubit observables, so we can check if it is possible to retrieve the original Peres-Mermin square for qubits within the formalism we developed in Section \ref{sec:operators}. 
Restricting to states with a single photon excitation per mode, we have a four dimensional subspace of a bipartite system (where the reduced state for each party has dimension two) and we recover the two qubits measurement scenario. As this subspace does not contain any diagonal states, we have the following:
\begin{equation}
 \langle\hat O\rangle_{qubit}=6.   
\end{equation}
We recall that it is the same quantum expectation value as if one would construct analog of (\ref{inOp}) using original Peres-Mermin square based on Pauli matrices \cite{Cabello:Contextuality}. Moreover, when the operators $\hat G_i^X$ act on states having one excitation per each beam $X$ (note that each beam consists of two bosonic modes), they behave exactly as the standard Pauli matrices. Therefore, the original result is retrieved with its state independence and the original Peres-Mermin paradox can be obtained in our approach. Note that state independence is also retrieved for subspaces of states restricted to odd number of bosons per beam as it does not contain any diagonal state. 

\section{Entanglement}
We address now if operators $\hat G_i$ can detect entanglement and whether they constitute advantage for this aim. 
To address the problem, we start proving that for any entanglement indicator written in the $n$-qubits of the form:
\begin{equation}
 \mathcal{\hat W}=\sum_{s_1,\dots,s_n=0}^3w_{s_1,\dots,s_n}\bigotimes_{i=1}^n\hat \sigma^i_{s_i},
\end{equation}
exists a correspondence that maps the $n-$qubits entanglement indicator into a $2n-$bosonic modes entanglement indicator, that reads: 
\begin{equation} \label{ISO}
\mathcal{\hat W}\rightarrow\mathcal{\hat W}_G =\sum_{s_1,\dots,s_n=0}^3w_{s_1,\dots,s_n}\prod_{i=1}^n\hat G^i_{s_i}.
\end{equation}
such that for any separable state $\rho_{sep}$ the following relation  holds 
\begin{equation}
\label{eq:sep_cond}
\Tr{\hat W_G \rho_{sep}}=\langle \mathcal{\hat W}_G\rangle_{sep}\geq 0.
\end{equation}
The proof proceeds in an analogous way to the one given in \cite{Stokes:mapping} for Stokes operators.
To show how the mapping works we start with the case of $n=2$. In order to prove (\ref{eq:sep_cond}) we have to show that for any mixed state  $\rho$ it is possible to find a density matrix  $\mathcal{\hat M}_{\rho}$  describing a two-qubit state, which fulfill the following relation:
\begin{equation}
\frac{\langle \mathcal{W}_G\rangle_{\rho}}{\langle\hat G_0^1\hat G_0^2\rangle_{\rho}}=\Tr[\mathcal{\hat W}\mathcal{\hat M}_{\rho}].\label{homrel}
\end{equation}
We compute the expectation value of $\hat G_\eta^1\hat G_\nu^2$ on a pure state $\ket{\phi}$ using for the operators the expressions in Eq.~(\ref{innaforma}), obtaining as a result:
\begin{align}
\begin{split}
\langle\hat G_\eta^1\hat G_\nu^2\rangle_{\ket{\phi}}&=\sum_{k,l=1}^2\sum_{n,m=1}^2\sigma_\eta^{kl}\sigma_\nu^{mn}\bra{\Phi_{km}}\ket{\Phi_{ln}}\\
&=\Tr[\hat \sigma^1_\eta\otimes\hat \sigma^2_\nu\hat M_{\phi}],
\end{split}
\end{align}
where $\ket{\Phi_{ln}}=\hat V_l^1\hat V_n^2\ket{\phi}$ and $\hat V_l^i$ is the $l$-th element of the vector $\overrightarrow{V}$ that acts on the modes of the $i$-th party as in (\ref{innaforma}). The matrix $\hat M_\phi$ has elements of the form $\bra{\Phi_{km}}\ket{\Phi_{ln}}$. This matrix is Gramian, therefore it is positive. Moreover, the following inequality:  
\begin{equation}
0<\Tr[\hat M_{\phi}]=\langle\hat G_0^1\hat G_0^2\rangle_{\phi}\leq 1,
\end{equation}
holds with the exception for cases where $\ket{\phi}$ is in the subspace of diagonal states. However, those cases are not interesting as the expectation value of $\mathcal{\hat W}_G$ is $0$ and thus it cannot change the bound for separable states (\ref{eq:sep_cond}). We can now define the density matrix for a two-qubits system which fulfills (\ref{homrel})  for a pure state $\rho=\ketbra{\phi}{\phi}$:
\begin{equation}
\mathcal{\hat M}_{\ket{\phi}}=\frac{\hat M_{\phi}}{\langle\hat G_0^1\hat G_0^2\rangle_{\phi}}.
\end{equation}
 For a mixed state $\rho$, which is a convex combination of states $\ket{\phi_i}\bra{\phi_i}$, we get the following density matrix:
\begin{equation}
\mathcal{\hat M}_{\rho}=\frac{\sum_i p_i\hat M_{\phi_i}}{\sum_i p_i\Tr[\hat M_{\phi_i}]}.\label{mrr}
\end{equation}

Consider a pure separable state of two-beam bosonic field $\ket{\psi^{12}}_{sep}=F_1^\dagger F_2^\dagger\ket{\Omega}$ where $F_X^\dagger$ is polynomial in the creation operators of the beam $X$. For such states, the following factorization occures:
\begin{equation}
_{sep}\bra{\Phi_{km}}\ket{\Phi_{ln}}_{sep}=\bra{\Psi_{k}^1}\ket{\Psi_{l}^1}\bra{\Psi_{m}^2}\ket{\Psi_{n}^2},
\end{equation} 
where $\ket{\Psi_{l}^X}=\hat V_l^X F_X^\dagger\ket{\Omega}$.
Thus, the matrix $\hat M_{\psi^{12}}$ factorizes into  $\hat M_{\psi}^1\hat M_{\psi}^2$, where matrix elements of $\hat M_{\psi}^X$ are given by $\bra{\Psi_{k}^X}\ket{\Psi_{l}^X}$. The matrix $\hat M_{\psi}^X$ after normalization is a well defined density matrix of qubit based on analogous arguments to those presented for $\mathcal{\hat M}_\rho$.
Due to the factorization of the density matrix $\mathcal{\hat M}_{\psi}$, it describes a pure separable state in the two-qubit space, thus:
\begin{equation}
\langle \mathcal{\hat W}\rangle_{\mathcal{\hat M}_{\psi}}\geq 0. \label{czym}
\end{equation}
As a mixed state is separable iff it is a convex combination of pure separable states, the matrix associated with such separable mixed state $\rho'$ of the form  (\ref{mrr}) is a convex combination of  two-qubit separable density matrices   $\hat M_{\psi_i}/\sum_i p_i\Tr[\hat M_{\psi_i}]$. Using  (\ref{czym}), we get that for any separable mixed state $\rho'$ following inequality holds:
\begin{equation}
\langle \mathcal{\hat W}\rangle_{\mathcal{\hat M}_{\rho'}}=\frac{\sum_i p_i\langle \mathcal{\hat W}\rangle_{\mathcal{\hat M}_{\psi_i}}}{\sum_i p_i\Tr[\hat M_{\psi_i}]}\geq 0.
\end{equation}
Thus, from the formula (\ref{homrel})  we get:
\begin{equation}
\langle \mathcal{\hat W}_G\rangle_{sep}=\langle \mathcal{\hat W}_G\rangle_{\rho '}\geq 0,
\end{equation}
which is the condition for $\mathcal{\hat W}_G$ to be an entanglement indicator.

We check if such a mapping is true for any $n$-qubit case. Let us start extending it to the three qubits scenario. In such a case, any pure partially separable state of the three-beam bosonic field $\ket{\psi^{123}}$ can be written in factorized form, for example $\ket{\psi^{123}}_{sep}=F_1^\dagger F_{23}^\dagger\ket{\Omega}$ where $F_{XY}^\dagger$ is a polynomial of creation operators associated with beams $X$ and $Y$.  Now we can analogously find the density matrix for $3$-qubit state  $\mathcal{\hat M}_{\psi^{123}}=\mathcal{\hat M}_{\psi}^{1}\mathcal{\hat M}_{\psi}^{23}$ which partially factorizes. Because the expectation value for any 3-qubit entanglement indicator is non-negative for partially factorizable 3-qubit density matrices, inequality (\ref{eq:sep_cond}) holds also for the three-beam case. Further extensions can be made following the previous arguments.

\subsection{Entanglement of the bright GHZ state}\label{BGHZent}
Now we can adopt entanglement indicators for $n$-qubit states to examine the entanglement of states of bosonic fields. Let us recall the bright GHZ state (BGHZ), whose construction and entanglement has been presented in \cite{BGHZ} for some range of amplification gain. This state is a generalization of two beam squeezed vacuum into the case of three beams. Current advancements in quantum optical experiments allow one to think about such generalizations to be feasible experimentally in the near future \cite{3phot_exp}. The usual methods used in the theoretical description of the generation of squeezed states, such as the parametric approximation turns out to be not well suited for this generalization. This is because the perturbation series in such approximation diverges. Still, using the Pad\'e approximation one can find approximate convergent parametric description. The resulting state has the following form:
\begin{align}
\begin{split}
\ket{BGHZ}=\sum_{k=0}^\infty\sum_{m=0}^k& C_{k-m}(\Gamma)C_m(\Gamma)\\
&\times(\hat a_1^\dagger\hat a_2^\dagger\hat a_3^\dagger )^{k-m}(\hat b_1^\dagger \hat b_2^\dagger \hat b_3^\dagger)^m\ket{\Omega}, 
\label{BGHZ0} 
\end{split}
\end{align}
where $\hat a_X^\dagger$, $\hat b_X^\dagger$ are creation operators in two orthogonal modes signed to an observer $X$ and $C_m(\Gamma)$ are complex coefficients which can be obtained with the method from \cite{BGHZ} for $\Gamma<0.9$. We can use our mapping
 $\sigma_i^1\otimes\sigma_j^2\otimes\sigma_k^3 \rightarrow \hat G_i^1\hat G_j^2\hat G_k^3$ to obtain a new entanglement indicator from the entanglement indicator tailored for GHZ state presented in \cite{GHZ:Indicator}:
\begin{align}
\begin{split}
\mathcal{W}=&\frac{3}{2}\hat G^1_0\hat G^2_0\hat G^3_0-\hat G_1^1\hat G_1^2\hat G_1^3\\
&- \frac{1}{2}(\hat G^1_0\hat G_3^2\hat G_3^3+\hat G_3^1\hat G^2_0\hat G_3^3+\hat G_3^1\hat G_3^2\hat G^3_0 ).\label{entBGHZ}
\end{split}
\end{align}
\begin{figure}[ht]
\centering
\includegraphics[width=0.49\textwidth]{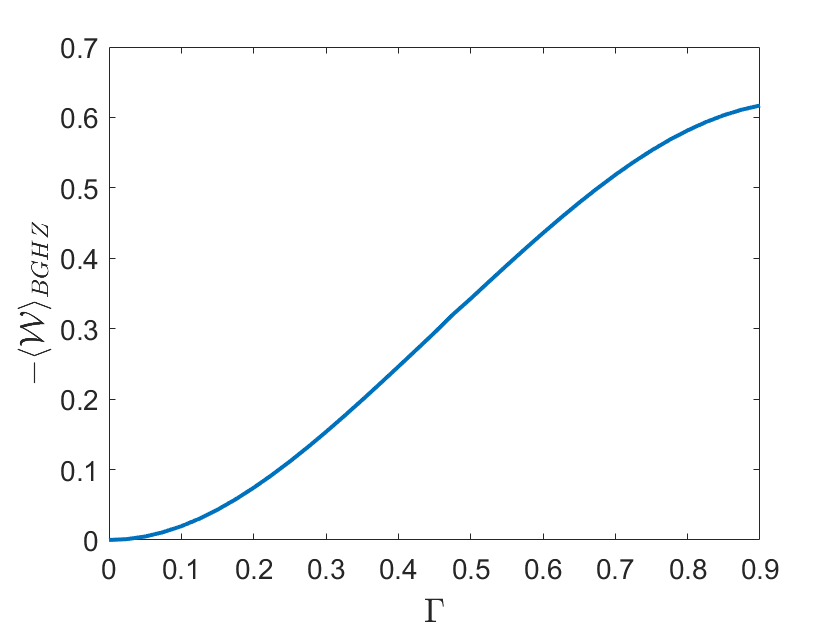} 
\caption{ Expectation values of the entanglement indicator $\mathcal{W}$  for the BGHZ state versus $\Gamma$ as in Equation \ref{entBGHZ}. From the curve one can observe that with increasing $\Gamma$ the probability of measuring eigenvalues associated with non-diagonal states increases. The range of $\Gamma$ is dictated by the fact that the method of calculating the probabilities of the BGHZ state has finite range of applicability \cite{BGHZ}. }
\label{entpa}
\end{figure}

We use now this indicator to check the entanglement of the BGHZ state.  $\ket{BGHZ}.$ 
It easy to see from Equation \ref{BGHZ0} that it can be written as a superposition of states:
\begin{multline}
\ket{\psi_{nm}}=\frac{1}{\sqrt{2}}\Big(\ket{n_{a_1},m_{b_1};n_{a_2}, m_{b_2};n_{a_3},m_{b_3}}\\+\ket{m_{a_1},n_{b_1};m_{a_2},n_{b_2};m_{a_3},n_{b_3}}\Big),\label{psinm}
\end{multline}
and diagonal states (with the same number of bosons in all modes) because states of superposition (\ref{psinm}) have the same amplitudes in (\ref{BGHZ0}) independently from the exact form of coefficients $C_q(\Gamma)$. Note that the states $\ket{\psi_{nm}}$ are eigenstates of the operator $\hat G_1^1\hat G_1^2\hat G_1^3$ with eigenvalue 1. This is because the first element of the superposition (\ref{psinm}) is turned into the second and vice versa  under the action of  $\hat G_1^1\hat G_1^2\hat G_1^3$. The action of such an operator on a diagonal state yields 0 and thus: 
\begin{equation}
\langle \hat G_1^1\hat G_1^2\hat G_1^3\rangle_{BGHZ}=\langle\hat G^1_0\hat G^2_0\hat G^3_0\rangle_{BGHZ}.\label{wo}
\end{equation} 

An analogous equality is true for the last three terms of (\ref{entBGHZ}) because those operators acting on $\ket{\psi_{nm}}$ give $(-1)^2$ and for diagonal states we get 0. As $\langle\hat G^1_0\hat G^2_0\hat G^3_0\rangle_{BGHZ}$ is simply the probability of observing result corresponding to non-diagonal states and it is always grater than $0$ for the BGHZ state for any $\Gamma>0$. As a consequence:
\begin{equation}
\langle \mathcal{W}\rangle_{BGHZ}=- \langle\hat G^1_0\hat G^2_0\hat G^3_0\rangle_{BGHZ}<0.
\end{equation}
From that follows that for any $\Gamma$ the state $\ket{BGHZ}$ is entangled. Fig. \ref{entpa} presents $-\langle \mathcal{W}\rangle_{BGHZ}$ versus $\Gamma$. We note that for terms of the BGHZ state with a fixed number of bosons per party $k$, the ratio between the number of diagonal states present in the superposition  and the number of  non-diagonal states drops to 0 also for even $k$. 
Moreover, because $\langle\hat G^1_0\hat G^2_0\hat G^3_0\rangle_{BGHZ}$ increases when $\Gamma$ increases, based on Fig. \ref{entBGHZ}, the probability of observing eigenvalues corresponding to diagonal states decreases. This is valid at least in the range where we performed calculations. Thus, the probability distribution  has no tendency to concentrate at the diagonal states with growing $\Gamma$ as having corresponding eigenvalue $0$ would indicate otherwise non-increasing pattern on the curve.
 However, as terms with greater photon numbers are generated by the same Hamiltonian such behaviour should be rather preserved also for $\Gamma>0.9$. Thus, the probability of obtaining a result corresponding to a diagonal state, namely $P(d|BGHZ)=1-\langle\hat G^1_0\hat G^2_0\hat G^3_0\rangle_{BGHZ},$ seems to decrease with growing $\Gamma$  for any of its values and converge to 0, as the number of terms whose amplitudes have a relevant magnitude  for the outcomes  grows, while the diagonal terms become a  minority among them and this effect is not countered by the concentration of probability distribution at the diagonal states. Thus, for $\Gamma\rightarrow\infty$ the expectation value of $\mathcal{W}$ should converge to the value of the 3-qubit case, namely -1. 

\subsection{Necessary and sufficient condition to detect entanglement for a state of $2\times 2$ modes  using operators $\hat G_i$}
A family of conditions for Pauli observables for two qubits  which together form a necessary and sufficient condition for entanglement has been proposed in \cite{Necessary}. It was adapted to the case of Stokes operators in \cite{Stokes:mapping} using a mapping that has inspired our work. Using the analogous mapping between qubit conditions and the studied case, shown in (\ref{ISO}), we can write a necessary and sufficient condition for {\it detecting} entanglement with the operators $\hat G_i$ for the two party scenario. 
It is 
\begin{align}
    \begin{split}
        \langle\hat G_1^1\hat G_1^2+\hat G_2^1\hat G_2^2\rangle^2&+\langle\hat G_3^1\hat G_0^2+\hat G_0^1\hat G_3^2\rangle^2\\
       & \leq\langle\hat G_0^1\hat G_0^2+\hat G_3^1\hat G_3^2\rangle^2.\label{CHINA_suff}
    \end{split}
\end{align}
and the family of conditions which can be obtained from (\ref{CHINA_suff}) by cyclically permuting the indides $1,2,3$ of one party or both.

We shall employ this criterion to study the entanglement of the BSV state as defined in Eq. (\ref{eq:bsv}). It is easy to notice that $\hat G_0^1\hat G_3^2\ket{BSV}=-\hat G_3^1\hat G_0^2\ket{BSV}$ and $\hat G_0^1\hat G_0^2\ket{BSV}=-\hat G_3^1\hat G_3^2\ket{BSV}$. Thus, we have to consider the first term of the LHS of the inequality (\ref{CHINA_suff}). One can also observe that $\hat G_1^1\hat G_1^2\ket{BSV}=\hat G_2^1\hat G_2^2\ket{BSV}$, because the sign part in $\hat G_2^1\hat G_2^2$ cancels the signs coming from $i^2$ when acting on the BSV state. 
It turns out that:
\begin{align}
    &\hat G_2^1\hat G_2^2\ket{\psi^{2n+1}}=-\ket{\psi^{2n+1}},\\
    &\hat G_2^1\hat G_2^2\ket{\psi^{2n}}=\hat G_0^1\hat G_0^2\ket{\psi^{2n}}.
\end{align}
We calculate the first term in LHS of (\ref{CHINA_suff}):
\begin{align}
  \begin{split}
      &\langle\hat G_1^1\hat G_1^2+\hat G_2^1\hat G_2^2\rangle^2=\\
      &4\left[\sum_{n=0}^\infty\left(\frac{\tanh^{4n}\Gamma}{\cosh^{4}\Gamma}-\frac{\tanh^{4n+2}\Gamma}{\cosh^{4}\Gamma}\right)-\sech 2\Gamma\right]^2\\
      &\;\;\;\;\;\;\;\;\;\;\;\;\;\;\;\;\;\;\;\;\;\;\;\;\;\;\;\;\;\;\;\;\;\;=16\sech^4(2\Gamma)\sinh^4\Gamma.
  \end{split}  
\end{align}
This expression is always grater than zero for $\Gamma\neq 0,$ therefore the operators $\hat G_i$ are able to detect entanglement for any $\Gamma>0$  in agreement to what has been showed using Stokes operators \cite{Stokes:mapping}. 

\section{Bell inequalities for the BGHZ state}
In this section, we would like to show that operators $\hat G_i$ can allow for detecting Bell nonclassicality. Let us introduce the modification of operators $\hat G_i$:
\begin{equation}
\hat G_{i-}=\hat G_i-\sum_{n=0}^\infty \ket{n,n}\bra{n,n}.
\end{equation}
This modification effectively assigns the value -1 to the cases when an equal number of bosons is measured in both modes. The introduced modification of the operators makes them dichotomic observables as the Pauli matrices with the same spectrum of $\pm 1$ . Therefore, we can write the Mermin inequality \cite{Mermin:inequality} using the operators $\hat G_{i-}^j$:  
\begin{align}
\begin{split}
&|\langle G_{1-}^1(\lambda) G_{1-}^2(\lambda) G_{1-}^3(\lambda)\\
&- G_{1-}^1(\lambda)G_{2-}^2(\lambda) G_{2-}^3(\lambda)
- G_{2-}^1(\lambda)G_{1-}^2(\lambda) G_{2-}^3(\lambda)\\
&-G_{2-}^1(\lambda)G_{2-}^2(\lambda) G_{1-}^3(\lambda)\rangle_{LHV}|\leq 2,
\end{split}
\label{BELLnowe}
\end{align}
where $\lambda$ is a local hidden variable and $G_{i-}^j(\lambda)$ are the local hidden values associated with the outcomes of measurement of operators $\hat G_{i-}^j$.

Now, we check whether it is possible to violate the inequality in (\ref{BELLnowe}) with the BGHZ state. As this state always has the same configuration of the number of bosons per any party, the operators
 $ G_{i-}^1 G_{j-}^2 G_{k-}^3$ effectively reduce to:

\begin{align}
\begin{split}
&\hat G_{i-}^1 \hat G_{j-}^2\hat G_{k-}^3\ket{BGHZ_3}
=\Big(\hat G_{i}^1 \hat G_{j}^2\hat G_{k}^3\\
&-\sum_{n=0}^\infty\ket{\mathbf{n}}\bra{\mathbf{n}}\Big) \ket{BGHZ_3}, 
\end{split}
\end{align}
where $\ket{\mathbf{n}}$ is the state having the same number of particles, $n$, in all the modes. 
Using (\ref{wo}) one can find that:
\begin{equation}
\langle\hat G_{1-}^1 \hat G_{1-}^2\hat G_{1-}^3\rangle_{BGHZ}=1-2P(d|BGHZ),
\end{equation}
where $P(d|BGHZ)=1-\langle\hat G^1_0\hat G^2_0\hat G^3_0\rangle_{BGHZ}$. We have used here the fact that only diagonal states contained in BGHZ are states with the same number of photons in all modes. One can also see that:
\begin{align}
\begin{split}
\langle\hat G_{1}^1 \hat G_{2}^2\hat G_{2}^3\rangle_{BGHZ}&=i^2\langle\hat G_{1}^1 \hat G_{1}^2\hat G_{1}^3\rangle_{BGHZ}\\
&=-\langle\hat G^1_0\hat G^2_0\hat G^3_0\rangle_{BGHZ}.
\end{split}
\end{align}
and from that follows:
\begin{align}
\begin{split}
\langle\hat G_{1-}^1 \hat G_{2-}^2\hat G_{2-}^3\rangle_{BGHZ}&=-\langle\hat G^1_0\hat G^2_0\hat G^3_0\rangle_{BGHZ}\\&-P(d|BGHZ)
=-1.
\end{split}
\end{align}
Analogous results hold for last two terms of (\ref{BELLnowe}), hence for the BGHZ state we get:
\begin{equation}
4-2P(d|BGHZ)\leq 2.\label{ghzbell}
\end{equation}
This inequality is violated for any $\Gamma>0$ because $2P(d|BGHZ)<2$ and as it was discussed in section \ref{BGHZent}, and for $\Gamma\rightarrow\infty$ we retrieve the 3-qubit case. This situation is opposite to the case where for constructing Mermin inequality normalized Stokes operators were used \cite{BGHZ}. This is in the sense that for normalized Stokes operators the range of violation is finite ($\Gamma<0.77$) and the strongest violation is obtained for low values of $\Gamma$.

The introduced modification allowed for violation in the whole range of $\Gamma$. Without this modification, one would get:
\begin{equation}
4-4P(d|BGHZ)\leq 2,   
\end{equation}
which is not violated for small $\Gamma$. However, in the limit of high values of $\Gamma,$ the behaviour of this inequality does not deviate significantly from (\ref{ghzbell}).

\section{Conclusions}
The quantification of nonclassical properties of bosonic fields has always played a pivotal role for the development of quantum technologies. Nowadays, the experimental techniques for counting the number of particles present in a bosonic mode  open new avenues to detect and quantify several nonclassical resources \cite{Walmsley17, Walmsley20, Stokes:original, NonCL_BEC}. However, special care must be taken when generalizing some concepts formulated for fixed number of particles, to situations inherently involving undefined particle numbers, see e.g. discussion in  \cite{1stPaper, 1stPLA, CommentDun, 3rdPaper}. 
Here, we introduce a set of new Stokes-like observables for two modes of a  bosonic field, which turns out to be a representation of $\mathfrak{su}(2)$ algebra. Using these operators we  construct a generalization of the Peres-Mermin square for a four-modes bosonic field, which allows to observe contextuality for states  with an undefined number of particles. This shows that contextuality is not only a phenomenon observable for
low dimensional quantum systems, and supersedes the approaches relying on the phase-space formalism \cite{Asadian:Contesxtuality, Contex_CV}. Particle-counting detection allows to extend the results to massive bosonic fields beyond the optical ones. Our generalization is not state independent, and the restriction to a single-boson
excitation per party retrieves the original formulation of the Peres-Mermin square. 

We also present  a mapping of $n$-qubit entanglement indicators to entanglement indicators for bosonic fields, which uses the Pauli-like obervables, which enter our Peres-Mermin square. This allows us e.g. to show that  a Bright GHZ state, a parametric process involving multiple emission to triple of polarization correlated photons, is entangled for any brightness. We also provide a necessary and sufficient  criterion for four mode states, which allows to detect entanglement with the Pauli-like observables. Its efficiency is shown for the $2\times 2$ bright squeezed vacuum state. Lastly, we  show a violation of the Bell inequality, based on the operators that we have introduced, for the Bright GHZ state in the whole range of the amplification gain. For all the three studied quantifiers of the nonclassicalicality of a field, we have observed that the results obtained for qubits can be recovered once one considers states with one particle in each beam heading to each detection station. Note that, the  previous generalizations of the Pauli formalism in an optical context, in form of standard Stokes operators, and normalized Stokes operators, see e.g. \cite{Zukowski:Stokes}, do not allow to generalize the Peres-Mermin construction.

As a final remark, we address an open question that arises from our results. How can be experimentally realized a measurement of such observables? Conceptually, the measurement of $\hat{G}_3,$ the last operator from proposed set, does not pose any difficulty, however measuring the another two operators seems to involve a highly nontrivial experimental procedure. This question may be of great significance for observing nonclassical aspects of quantum optical fields for macroscopic states as those observables seem to be optimal for this kind of states. Thus, this construction can stimulate advancements in experimental methods in quantum optics.
Once it will be achieved, we forecast that such a strategy can also be fruitful to quantify the nonclassicality of other bosonic fields, such as many-body systems of massive particle composing Bose Einstein condensate \cite{NonCL_BEC}. Another interesting task is to operationally construct a generalization of the proposed scheme to also cover optical realizations of generalized Gell-Mann observables. This would involve multiport beamsplitters \cite{multiport}.

\section*{Acknowledgements}
This work is supported by  Foundation for Polish Science (FNP), IRAP project ICTQT, contract no. 2018/MAB/5, co-financed by EU  Smart Growth Operational Programme.
AM is supported by (Polish) National Science Center (NCN): MINIATURA  DEC-2020/04/X/ST2/01794.

\appendix 
\section{Anticomutation and commutation relations}\label{comm}
\setcounter{equation}{0}
Firstly, let us check if $\lbrace\hat G_1,\hat G_3\rbrace=0$. This can be verified by acting with $\hat G_1\hat G_3$ and $\hat G_3\hat G_1$ on the basis vectors $\ket{n,m}$ and $\ket{m,n}$ for $n>m$. Knowing that $\hat G_1$ swaps modes and that states $\ket{n,m}$($\ket{m,n}$) are eigenvectors of  $\hat G_3$ with eigenvalue 1(-1) we get for $\ket{n,m}$:   
\begin{align*}
&\hat G_1\hat G_3 \ket{n,m}=\hat G_1\ket{n,m}=\ket{m,n},\\
&\hat G_3\hat G_1 \ket{n,m}=\hat G_3\ket{m,n}=-\ket{m,n},
\end{align*}
and for $\ket{mn}$
\begin{align*}
&\hat G_1\hat G_3 \ket{mn}=-\hat G_1\ket{mn}=-\ket{nm},\\
&\hat G_3\hat G_1 \ket{mn}=\hat G_3\ket{nm}=\ket{nm}.
\end{align*}
Because diagonal states have eigenvalue 0 for both those operators, we have $\hat G_3\hat G_1=-\hat G_1\hat G_3$ and thus, the anticommutation relation is fulfilled.

Now let us check if $\lbrace\hat G_1,\hat G_2\rbrace=0$. One can observe that $  \hat G_2=-i\hat G_3\hat G_1$ and so we have:
\begin{align*}
\hat G_2\hat G_1&=-i\hat G_3\hat G_1 \hat G_1=-i\hat G_3\hat G_0=-i\hat G_3,\\
\begin{split}
\hat G_1\hat G_2&=-i\hat G_1\hat G_3 \hat G_1=-i\hat G_1(-\hat G_1\hat G_3)\\
&=i\hat G_0\hat G_3=i\hat G_0\hat G_3.
\end{split}
\end{align*}
In the previous expression, we have used the fact that $\hat G_1^2=\hat G_0$ which is easily seen form the fact that flipping modes two times does not change a state. We have also used that $\hat G_0\hat G_i=\hat G_i\hat G_0=\hat G_i$ which follows from that $\hat G_0$ has a common zero eigenvalue for diagonal states with any $\hat G_i$ and it has eigenvalue 1 for other states (this  also shows that $\hat G_0$ commutes with any $\hat G_i$). As a result, we obtain $\lbrace\hat G_1,\hat G_2\rbrace=0$  

Now we check if $\lbrace\hat G_2,\hat G_3\rbrace=0$. Using the fact that $\hat G_3^2=\hat G_0$ we get:
\begin{align*}
&\hat G_2\hat G_3=-i\hat G_3\hat G_1 \hat G_3=-i\hat G_3(-\hat G_3\hat G_1)=i\hat G_0\hat G_1,\\
&\hat G_3\hat G_2=-i\hat G_3\hat G_3 \hat G_1=-i\hat G_0\hat G_1=-i\hat G_1.
\end{align*}
Thus, the anticommutation relation under consideration holds.

Lastly, we check if $\hat G_2^2=\hat G_0$:
\begin{align*}
\hat G_2^2&=-i\hat G_3\hat G_1(-i\hat G_3\hat G_1)=-\hat G_3\hat G_1\hat G_3\hat G_1\\
&=\hat G_3\hat G_3\hat G_1\hat G_1=\hat G_0^2=\hat G_0.
\end{align*}

Taking all these relations together, we have:
\begin{equation*}
\lbrace\hat G_i,\hat G_j\rbrace=2\delta_{ij}\hat G_0.
\end{equation*}

Let us now check the commutation relations. We start with $[ \hat G_3,\hat G_1]$. Using the anticomutation relations, previously proved, we have:
\begin{align}
\hat G_3\hat G_1-\hat G_1\hat G_3=2\hat G_3\hat G_1=2i\hat G_2,
\end{align}
where the last equality follows from $  \hat G_2=-i\hat G_3\hat G_1$. 
In case of $[ \hat G_1,\hat G_2]$ we have:
\begin{align}
\begin{split}
\hat G_1\hat G_2-\hat G_2\hat G_1&=-2\hat G_2\hat G_1\\
&=2i\hat G_3\hat G_1\hat G_1=2i\hat G_3.
\end{split}
\end{align}
Finally, for $[ \hat G_2,\hat G_3]$ we have:
\begin{align}
\begin{split}
\hat G_2\hat G_3-\hat G_3\hat G_2&=-2\hat G_3\hat G_2\\
&=2i\hat G_3\hat G_3\hat G_1=2i\hat G_1.
\end{split}
\end{align}
Thus, the commutation relations for the set of operators $\hat G_i$ are as follows:
\begin{equation}
[ \hat G_i,\hat G_j]=2i\epsilon_{ijk}\hat G_k
\end{equation}
Adding the anticommutation relation and commutation relation, we obtain:
\begin{align*}
&(\hat G_i\hat G_j-\hat G_j\hat G_i)+(\hat G_i\hat G_j+\hat G_j\hat G_i)\\
&=2\hat G_i\hat G_j=2i\epsilon_{ijk}\hat G_k+2\delta_{ij}\hat G_0,
\end{align*}
and from this follows an analogous relation as for the Pauli matrices:
\begin{equation}
\hat G_i\hat G_j=\delta_{ij}\hat G_0+i\epsilon_{ijk}\hat G_k.
\end{equation}
\section{Spectrum of operators $\hat G_i$}\label{spectrum}
We can characterize the set of eigenvectors for operators $\hat G_i$. In the case of $\hat G_1$ we have:
\begin{align*}
&\frac{1}{\sqrt{2}}(\ket{n,m}+\ket{m,n})\;\;\;\;\;\; \textrm{for eigenvalue 1},\\
&\frac{1}{\sqrt{2}}(\ket{n,m}-\ket{m,n})\;\;\;\;\;\; \textrm{for eigenvalue -1},\\
&\ket{nn}\;\;\;\;\;\;\;\;\;\;\;\;\;\;\;\;\;\;\;\;\;\;\;\;\;\; \textrm{for eigenvalue 0}.\\
\end{align*}
Let us now assume that $n>m$. For the operator $\hat G_2$ we obtain:
\begin{align*}
&\frac{1}{\sqrt{2}}(\ket{n,m}+i\ket{m,n})\;\;\;\;\;\; \textrm{for eigenvalue 1} \\
&\frac{1}{\sqrt{2}}(\ket{n,m}-i\ket{m,n})\;\;\;\;\;\; \textrm{for eigenvalue -1 },\\
&\ket{n,n}\;\;\;\;\;\;\;\;\;\;\;\;\;\;\;\;\;\;\;\;\;\;\;\;\;\;\;\; \textrm{for eigenvalue 0}.\\
\end{align*}
Finally, for the operator $\hat G_3$ we have:
\begin{align*}
&\ket{n,m}\;\;\;\;\;\;\;\;\;\;\;\;\;\;\;\;\;\;\; \textrm{for eigenvalue 1 },\\
&\ket{m,n}\;\;\;\;\;\;\;\;\;\;\;\;\;\;\;\;\;\;\; \textrm{for eigenvalue -1 },\\
&\ket{n,n}\;\;\;\;\;\;\;\;\;\;\;\;\;\;\;\;\;\;\;\; \textrm{for eigenvalue 0}.\\
\end{align*}
Considering the three basis vectors $\ket{n,n},\ket{n,m},\ket{m,n}$ we can write the three orthonormal eigenvectors for any of the operators, we can sum up that this set of eigenvectors characterizes all eigenvectors of $\hat G_i$ and that the spectrum of operators $\hat G_i$ is $\pm 1,0$.

\section{Unitary transformation of modes}\label{ap:rot}
\setcounter{equation}{0}
It can be shown that the set of operators $\hat S_i=\frac12 A^\dagger\sigma_i A$ after unitary transformation,
which is equivalent to cyclic permutation of the basis, 
will result in an equivalent set of operators only changing their labelling.  However, the same is not true for the set of operators $\hat G_i$

To see that, let us consider two orthonormal mutually unbiased polarization bases $\lbrace H,V\rbrace$ (horizontal,vertical) and $\lbrace D,A\rbrace$ (diagonal, anti-diagonal) and their creation operators, respectively: $a_H^\dagger,\,b_V^\dagger$ and $a_D^\dagger\,b_A^\dagger$. If one writes the $\hat S_3^{(D)}$ operator based on $a_D^\dagger,b_A^\dagger$ one expects that this operator rewritten in the $\lbrace H,V\rbrace$ basis will coincide with the operator $\hat S_1^{(H)}$ based on $a_H^\dagger,b_V^\dagger$. We show that this is not true for operators $\hat G_i$ on the basis of a counterexample.

As $\hat G_i$ and $\hat S_i$ coincide in the one photon subspace up to coefficient, the first nontrivial counterexample can be shown  in the two-photon subspace. Let us consider $\hat G_3^{(D)}$ restricted to two photon subspace:
\begin{equation}
  \hat G_3^{(D|2)}:=\ket{2_D,0_A}\bra{2_D,0_A}  -\ket{0_D,2_A}\bra{0_D,2_A}.
\end{equation}
After rewriting this operator in $\lbrace H,V\rbrace$ basis we obtain:
\begin{multline}
  \hat G_3^{(D|2)}=-\frac{1}{\sqrt{2}}\Big(\ket{2_H,0_V}\bra{1_H,1_V}+\ket{0_H,2_V}\bra{1_H,1_V}\\+\ket{1_H,1_V}\bra{2_H,0_V}+\ket{1_H,1_V}\bra{0_H,2_V}  \Big) .
\end{multline}
As this operator does not coincide with $\hat G_1^{(H|2)}$, the operators $\hat G_i$ do not share the considered property with the operators $\hat S_i$.


\begin{thebibliography}{33}%
\makeatletter
\providecommand \@ifxundefined [1]{%
 \@ifx{#1\undefined}
}%
\providecommand \@ifnum [1]{%
 \ifnum #1\expandafter \@firstoftwo
 \else \expandafter \@secondoftwo
 \fi
}%
\providecommand \@ifx [1]{%
 \ifx #1\expandafter \@firstoftwo
 \else \expandafter \@secondoftwo
 \fi
}%
\providecommand \natexlab [1]{#1}%
\providecommand \enquote  [1]{``#1''}%
\providecommand \bibnamefont  [1]{#1}%
\providecommand \bibfnamefont [1]{#1}%
\providecommand \citenamefont [1]{#1}%
\providecommand \href@noop [0]{\@secondoftwo}%
\providecommand \href [0]{\begingroup \@sanitize@url \@href}%
\providecommand \@href[1]{\@@startlink{#1}\@@href}%
\providecommand \@@href[1]{\endgroup#1\@@endlink}%
\providecommand \@sanitize@url [0]{\catcode `\\12\catcode `\$12\catcode
  `\&12\catcode `\#12\catcode `\^12\catcode `\_12\catcode `\%12\relax}%
\providecommand \@@startlink[1]{}%
\providecommand \@@endlink[0]{}%
\providecommand \url  [0]{\begingroup\@sanitize@url \@url }%
\providecommand \@url [1]{\endgroup\@href {#1}{\urlprefix }}%
\providecommand \urlprefix  [0]{URL }%
\providecommand \Eprint [0]{\href }%
\providecommand \doibase [0]{http://dx.doi.org/}%
\providecommand \selectlanguage [0]{\@gobble}%
\providecommand \bibinfo  [0]{\@secondoftwo}%
\providecommand \bibfield  [0]{\@secondoftwo}%
\providecommand \translation [1]{[#1]}%
\providecommand \BibitemOpen [0]{}%
\providecommand \bibitemStop [0]{}%
\providecommand \bibitemNoStop [0]{.\EOS\space}%
\providecommand \EOS [0]{\spacefactor3000\relax}%
\providecommand \BibitemShut  [1]{\csname bibitem#1\endcsname}%
\let\auto@bib@innerbib\@empty
\bibitem [{\citenamefont {Howard}\ \emph {et~al.}(2014)\citenamefont {Howard},
  \citenamefont {Wallman}, \citenamefont {Veitch},\ and\ \citenamefont
  {Emerson}}]{ContextualityAdv}%
  \BibitemOpen
  \bibfield  {author} {\bibinfo {author} {\bibfnamefont {Mark}\ \bibnamefont
  {Howard}}, \bibinfo {author} {\bibfnamefont {Joel}\ \bibnamefont {Wallman}},
  \bibinfo {author} {\bibfnamefont {Victor}\ \bibnamefont {Veitch}}, \ and\
  \bibinfo {author} {\bibfnamefont {Joseph}\ \bibnamefont {Emerson}},\
  }\bibfield  {title} {\enquote {\bibinfo {title} {Contextuality supplies the
  “magic” for quantum computation},}\ }\href {\doibase 10.1038/nature13460}
  {\bibfield  {journal} {\bibinfo  {journal} {Nature}\ }\textbf {\bibinfo
  {volume} {510}},\ \bibinfo {pages} {351–355} (\bibinfo {year}
  {2014})}\BibitemShut {NoStop}%
\bibitem [{\citenamefont {Vazirani}\ and\ \citenamefont
  {Vidick}(2014)}]{DeviceindepQKD}%
  \BibitemOpen
  \bibfield  {author} {\bibinfo {author} {\bibfnamefont {Umesh}\ \bibnamefont
  {Vazirani}}\ and\ \bibinfo {author} {\bibfnamefont {Thomas}\ \bibnamefont
  {Vidick}},\ }\bibfield  {title} {\enquote {\bibinfo {title} {Fully
  device-independent quantum key distribution},}\ }\href {\doibase
  10.1103/PhysRevLett.113.140501} {\bibfield  {journal} {\bibinfo  {journal}
  {Phys. Rev. Lett.}\ }\textbf {\bibinfo {volume} {113}},\ \bibinfo {pages}
  {140501} (\bibinfo {year} {2014})}\BibitemShut {NoStop}%
\bibitem [{\citenamefont {Giovannetti}\ \emph {et~al.}(2004)\citenamefont
  {Giovannetti}, \citenamefont {Lloyd},\ and\ \citenamefont
  {Maccone}}]{Giovannetti_metr}%
  \BibitemOpen
  \bibfield  {author} {\bibinfo {author} {\bibfnamefont {Vittorio}\
  \bibnamefont {Giovannetti}}, \bibinfo {author} {\bibfnamefont {Seth}\
  \bibnamefont {Lloyd}}, \ and\ \bibinfo {author} {\bibfnamefont {Lorenzo}\
  \bibnamefont {Maccone}},\ }\bibfield  {title} {\enquote {\bibinfo {title}
  {Quantum-enhanced measurements: Beating the standard quantum limit},}\ }\href
  {\doibase 10.1126/science.1104149} {\bibfield  {journal} {\bibinfo  {journal}
  {Science}\ }\textbf {\bibinfo {volume} {306}},\ \bibinfo {pages}
  {1330–1336} (\bibinfo {year} {2004})}\BibitemShut {NoStop}%
\bibitem [{\citenamefont {Mandel}\ and\ \citenamefont
  {Wolf}(1995)}]{mandel_wolf_1995}%
  \BibitemOpen
  \bibfield  {author} {\bibinfo {author} {\bibfnamefont {Leonard}\ \bibnamefont
  {Mandel}}\ and\ \bibinfo {author} {\bibfnamefont {Emil}\ \bibnamefont
  {Wolf}},\ }\href {\doibase 10.1017/CBO9781139644105} {\emph {\bibinfo {title}
  {Optical Coherence and Quantum Optics}}}\ (\bibinfo  {publisher} {Cambridge
  University Press},\ \bibinfo {year} {1995})\BibitemShut {NoStop}%
\bibitem [{\citenamefont {Pan}\ \emph {et~al.}(2012)\citenamefont {Pan},
  \citenamefont {Chen}, \citenamefont {Lu}, \citenamefont {Weinfurter},
  \citenamefont {Zeilinger},\ and\ \citenamefont {\ifmmode~\dot{Z}\else
  \.{Z}\fi{}ukowski}}]{Pan2012}%
  \BibitemOpen
  \bibfield  {author} {\bibinfo {author} {\bibfnamefont {Jian-Wei}\
  \bibnamefont {Pan}}, \bibinfo {author} {\bibfnamefont {Zeng-Bing}\
  \bibnamefont {Chen}}, \bibinfo {author} {\bibfnamefont {Chao-Yang}\
  \bibnamefont {Lu}}, \bibinfo {author} {\bibfnamefont {Harald}\ \bibnamefont
  {Weinfurter}}, \bibinfo {author} {\bibfnamefont {Anton}\ \bibnamefont
  {Zeilinger}}, \ and\ \bibinfo {author} {\bibfnamefont {Marek}\ \bibnamefont
  {\ifmmode~\dot{Z}\else \.{Z}\fi{}ukowski}},\ }\bibfield  {title} {\enquote
  {\bibinfo {title} {Multiphoton entanglement and interferometry},}\ }\href
  {\doibase 10.1103/RevModPhys.84.777} {\bibfield  {journal} {\bibinfo
  {journal} {Rev. Mod. Phys.}\ }\textbf {\bibinfo {volume} {84}},\ \bibinfo
  {pages} {777--838} (\bibinfo {year} {2012})}\BibitemShut {NoStop}%
\bibitem [{\citenamefont {Mermin}(1993)}]{Mermin1993}%
  \BibitemOpen
  \bibfield  {author} {\bibinfo {author} {\bibfnamefont {N.~David}\
  \bibnamefont {Mermin}},\ }\bibfield  {title} {\enquote {\bibinfo {title}
  {Hidden variables and the two theorems of {John} {Bell}},}\ }\href {\doibase
  10.1103/RevModPhys.65.803} {\bibfield  {journal} {\bibinfo  {journal} {Rev.
  Mod. Phys.}\ }\textbf {\bibinfo {volume} {65}},\ \bibinfo {pages} {803--815}
  (\bibinfo {year} {1993})}\BibitemShut {NoStop}%
\bibitem [{\citenamefont {Schlichtholz}\ \emph
  {et~al.}(2021{\natexlab{a}})\citenamefont {Schlichtholz}, \citenamefont
  {Woloncewicz},\ and\ \citenamefont {Żukowski}}]{schlichtholz2021simplified}%
  \BibitemOpen
  \bibfield  {author} {\bibinfo {author} {\bibfnamefont {K.}~\bibnamefont
  {Schlichtholz}}, \bibinfo {author} {\bibfnamefont {B.}~\bibnamefont
  {Woloncewicz}}, \ and\ \bibinfo {author} {\bibfnamefont {M.}~\bibnamefont
  {Żukowski}},\ }\href@noop {} {\enquote {\bibinfo {title} {Simplified quantum
  optical stokes observables and {Bell}'s theorem},}\ } (\bibinfo {year}
  {2021}{\natexlab{a}}),\ \Eprint {http://arxiv.org/abs/2112.00084}
  {arXiv:2112.00084 [quant-ph]} \BibitemShut {NoStop}%
\bibitem [{\citenamefont {Wodkiewicz}\ and\ \citenamefont
  {Eberly}(1985)}]{algebra_su2}%
  \BibitemOpen
  \bibfield  {author} {\bibinfo {author} {\bibfnamefont {K}~\bibnamefont
  {Wodkiewicz}}\ and\ \bibinfo {author} {\bibfnamefont {JH}~\bibnamefont
  {Eberly}},\ }\bibfield  {title} {\enquote {\bibinfo {title} {Coherent states,
  squeezed fluctuations, and the su (2) am su (1, 1) groups in quantum-optics
  applications},}\ }\href@noop {} {\bibfield  {journal} {\bibinfo  {journal}
  {JOSA B}\ }\textbf {\bibinfo {volume} {2}},\ \bibinfo {pages} {458--466}
  (\bibinfo {year} {1985})}\BibitemShut {NoStop}%
\bibitem [{\citenamefont {Kochen}\ and\ \citenamefont {Specker}(1967)}]{KS_th}%
  \BibitemOpen
  \bibfield  {author} {\bibinfo {author} {\bibfnamefont {S}~\bibnamefont
  {Kochen}}\ and\ \bibinfo {author} {\bibfnamefont {EP}~\bibnamefont
  {Specker}},\ }\bibfield  {title} {\enquote {\bibinfo {title} {The problem of
  hidden variables in quantum mechanics},}\ }\href@noop {} {\bibfield
  {journal} {\bibinfo  {journal} {J. Math. Mech.}\ }\textbf {\bibinfo {volume}
  {17}},\ \bibinfo {pages} {59--87} (\bibinfo {year} {1967})}\BibitemShut
  {NoStop}%
\bibitem [{\citenamefont {Michler}\ \emph {et~al.}(2000)\citenamefont
  {Michler}, \citenamefont {Weinfurter},\ and\ \citenamefont
  {\ifmmode~\dot{Z}\else \.{Z}\fi{}ukowski}}]{Context_exp1}%
  \BibitemOpen
  \bibfield  {author} {\bibinfo {author} {\bibfnamefont {Markus}\ \bibnamefont
  {Michler}}, \bibinfo {author} {\bibfnamefont {Harald}\ \bibnamefont
  {Weinfurter}}, \ and\ \bibinfo {author} {\bibfnamefont {Marek}\ \bibnamefont
  {\ifmmode~\dot{Z}\else \.{Z}\fi{}ukowski}},\ }\bibfield  {title} {\enquote
  {\bibinfo {title} {Experiments towards falsification of noncontextual hidden
  variable theories},}\ }\href {\doibase 10.1103/PhysRevLett.84.5457}
  {\bibfield  {journal} {\bibinfo  {journal} {Phys. Rev. Lett.}\ }\textbf
  {\bibinfo {volume} {84}},\ \bibinfo {pages} {5457--5461} (\bibinfo {year}
  {2000})}\BibitemShut {NoStop}%
\bibitem [{\citenamefont {Lapkiewicz}\ \emph {et~al.}(2011)\citenamefont
  {Lapkiewicz}, \citenamefont {Li}, \citenamefont {Schaeff}, \citenamefont
  {Langford}, \citenamefont {Ramelow}, \citenamefont {Wie{\'s}niak},\ and\
  \citenamefont {Zeilinger}}]{Context_exp2}%
  \BibitemOpen
  \bibfield  {author} {\bibinfo {author} {\bibfnamefont {Radek}\ \bibnamefont
  {Lapkiewicz}}, \bibinfo {author} {\bibfnamefont {Peizhe}\ \bibnamefont {Li}},
  \bibinfo {author} {\bibfnamefont {Christoph}\ \bibnamefont {Schaeff}},
  \bibinfo {author} {\bibfnamefont {Nathan~K}\ \bibnamefont {Langford}},
  \bibinfo {author} {\bibfnamefont {Sven}\ \bibnamefont {Ramelow}}, \bibinfo
  {author} {\bibfnamefont {Marcin}\ \bibnamefont {Wie{\'s}niak}}, \ and\
  \bibinfo {author} {\bibfnamefont {Anton}\ \bibnamefont {Zeilinger}},\
  }\bibfield  {title} {\enquote {\bibinfo {title} {Experimental
  non-classicality of an indivisible quantum system},}\ }\href@noop {}
  {\bibfield  {journal} {\bibinfo  {journal} {Nature}\ }\textbf {\bibinfo
  {volume} {474}},\ \bibinfo {pages} {490--493} (\bibinfo {year}
  {2011})}\BibitemShut {NoStop}%
\bibitem [{\citenamefont {Asadian}\ \emph {et~al.}(2015)\citenamefont
  {Asadian}, \citenamefont {Budroni}, \citenamefont {Steinhoff}, \citenamefont
  {Rabl},\ and\ \citenamefont {G\"uhne}}]{Asadian:Contesxtuality}%
  \BibitemOpen
  \bibfield  {author} {\bibinfo {author} {\bibfnamefont {Ali}\ \bibnamefont
  {Asadian}}, \bibinfo {author} {\bibfnamefont {Costantino}\ \bibnamefont
  {Budroni}}, \bibinfo {author} {\bibfnamefont {Frank E.~S.}\ \bibnamefont
  {Steinhoff}}, \bibinfo {author} {\bibfnamefont {Peter}\ \bibnamefont {Rabl}},
  \ and\ \bibinfo {author} {\bibfnamefont {Otfried}\ \bibnamefont {G\"uhne}},\
  }\bibfield  {title} {\enquote {\bibinfo {title} {Contextuality in phase
  space},}\ }\href {\doibase 10.1103/PhysRevLett.114.250403} {\bibfield
  {journal} {\bibinfo  {journal} {Phys. Rev. Lett.}\ }\textbf {\bibinfo
  {volume} {114}},\ \bibinfo {pages} {250403} (\bibinfo {year}
  {2015})}\BibitemShut {NoStop}%
\bibitem [{\citenamefont {McKeown}\ \emph {et~al.}(2011)\citenamefont
  {McKeown}, \citenamefont {Paris},\ and\ \citenamefont
  {Paternostro}}]{Contex_CV}%
  \BibitemOpen
  \bibfield  {author} {\bibinfo {author} {\bibfnamefont {Gerard}\ \bibnamefont
  {McKeown}}, \bibinfo {author} {\bibfnamefont {Matteo G.~A.}\ \bibnamefont
  {Paris}}, \ and\ \bibinfo {author} {\bibfnamefont {Mauro}\ \bibnamefont
  {Paternostro}},\ }\bibfield  {title} {\enquote {\bibinfo {title} {Testing
  quantum contextuality of continuous-variable states},}\ }\href {\doibase
  10.1103/PhysRevA.83.062105} {\bibfield  {journal} {\bibinfo  {journal} {Phys.
  Rev. A}\ }\textbf {\bibinfo {volume} {83}},\ \bibinfo {pages} {062105}
  (\bibinfo {year} {2011})}\BibitemShut {NoStop}%
\bibitem [{\citenamefont {Peres}(1990)}]{PERES:PM}%
  \BibitemOpen
  \bibfield  {author} {\bibinfo {author} {\bibfnamefont {Asher}\ \bibnamefont
  {Peres}},\ }\bibfield  {title} {\enquote {\bibinfo {title} {Incompatible
  results of quantum measurements},}\ }\href {\doibase
  https://doi.org/10.1016/0375-9601(90)90172-K} {\bibfield  {journal} {\bibinfo
   {journal} {Physics Letters A}\ }\textbf {\bibinfo {volume} {151}},\ \bibinfo
  {pages} {107--108} (\bibinfo {year} {1990})}\BibitemShut {NoStop}%
\bibitem [{\citenamefont {Mermin}(1990{\natexlab{a}})}]{Mermin:PM}%
  \BibitemOpen
  \bibfield  {author} {\bibinfo {author} {\bibfnamefont {N.~David}\
  \bibnamefont {Mermin}},\ }\bibfield  {title} {\enquote {\bibinfo {title}
  {Simple unified form for the major no-hidden-variables theorems},}\ }\href
  {\doibase 10.1103/PhysRevLett.65.3373} {\bibfield  {journal} {\bibinfo
  {journal} {Phys. Rev. Lett.}\ }\textbf {\bibinfo {volume} {65}},\ \bibinfo
  {pages} {3373--3376} (\bibinfo {year} {1990}{\natexlab{a}})}\BibitemShut
  {NoStop}%
\bibitem [{\citenamefont {Cabello}(2008)}]{Cabello:Contextuality}%
  \BibitemOpen
  \bibfield  {author} {\bibinfo {author} {\bibfnamefont {Ad\'an}\ \bibnamefont
  {Cabello}},\ }\bibfield  {title} {\enquote {\bibinfo {title} {Experimentally
  testable state-independent quantum contextuality},}\ }\href {\doibase
  10.1103/PhysRevLett.101.210401} {\bibfield  {journal} {\bibinfo  {journal}
  {Phys. Rev. Lett.}\ }\textbf {\bibinfo {volume} {101}},\ \bibinfo {pages}
  {210401} (\bibinfo {year} {2008})}\BibitemShut {NoStop}%
\bibitem [{\citenamefont {\ifmmode~\dot{Z}\else \.{Z}\fi{}ukowski}\ \emph
  {et~al.}(2017)\citenamefont {\ifmmode~\dot{Z}\else \.{Z}\fi{}ukowski},
  \citenamefont {Laskowski},\ and\ \citenamefont {Wie\ifmmode~\acute{s}\else
  \'{s}\fi{}niak}}]{ZUKUPRA}%
  \BibitemOpen
  \bibfield  {author} {\bibinfo {author} {\bibfnamefont {Marek}\ \bibnamefont
  {\ifmmode~\dot{Z}\else \.{Z}\fi{}ukowski}}, \bibinfo {author} {\bibfnamefont
  {Wies\l{}aw}\ \bibnamefont {Laskowski}}, \ and\ \bibinfo {author}
  {\bibfnamefont {Marcin}\ \bibnamefont {Wie\ifmmode~\acute{s}\else
  \'{s}\fi{}niak}},\ }\bibfield  {title} {\enquote {\bibinfo {title}
  {Normalized {Stokes} operators for polarization correlations of entangled
  optical fields},}\ }\href {\doibase 10.1103/PhysRevA.95.042113} {\bibfield
  {journal} {\bibinfo  {journal} {Phys. Rev. A}\ }\textbf {\bibinfo {volume}
  {95}},\ \bibinfo {pages} {042113} (\bibinfo {year} {2017})}\BibitemShut
  {NoStop}%
\bibitem [{\citenamefont {Ryu}\ \emph {et~al.}(2019)\citenamefont {Ryu},
  \citenamefont {Woloncewicz}, \citenamefont {Marciniak}, \citenamefont
  {Wie\ifmmode~\acute{s}\else \'{s}\fi{}niak},\ and\ \citenamefont
  {\ifmmode~\dot{Z}\else \.{Z}\fi{}ukowski}}]{Stokes:mapping}%
  \BibitemOpen
  \bibfield  {author} {\bibinfo {author} {\bibfnamefont {Junghee}\ \bibnamefont
  {Ryu}}, \bibinfo {author} {\bibfnamefont {Bianka}\ \bibnamefont
  {Woloncewicz}}, \bibinfo {author} {\bibfnamefont {Marcin}\ \bibnamefont
  {Marciniak}}, \bibinfo {author} {\bibfnamefont {Marcin}\ \bibnamefont
  {Wie\ifmmode~\acute{s}\else \'{s}\fi{}niak}}, \ and\ \bibinfo {author}
  {\bibfnamefont {Marek}\ \bibnamefont {\ifmmode~\dot{Z}\else
  \.{Z}\fi{}ukowski}},\ }\bibfield  {title} {\enquote {\bibinfo {title}
  {General mapping of multiqudit entanglement conditions to nonseparability
  indicators for quantum-optical fields},}\ }\href {\doibase
  10.1103/PhysRevResearch.1.032041} {\bibfield  {journal} {\bibinfo  {journal}
  {Phys. Rev. Research}\ }\textbf {\bibinfo {volume} {1}},\ \bibinfo {pages}
  {032041} (\bibinfo {year} {2019})}\BibitemShut {NoStop}%
\bibitem [{\citenamefont {Schlichtholz}\ \emph
  {et~al.}(2021{\natexlab{b}})\citenamefont {Schlichtholz}, \citenamefont
  {Woloncewicz},\ and\ \citenamefont {\ifmmode~\dot{Z}\else
  \.{Z}\fi{}ukowski}}]{BGHZ}%
  \BibitemOpen
  \bibfield  {author} {\bibinfo {author} {\bibfnamefont {Konrad}\ \bibnamefont
  {Schlichtholz}}, \bibinfo {author} {\bibfnamefont {Bianka}\ \bibnamefont
  {Woloncewicz}}, \ and\ \bibinfo {author} {\bibfnamefont {Marek}\ \bibnamefont
  {\ifmmode~\dot{Z}\else \.{Z}\fi{}ukowski}},\ }\bibfield  {title} {\enquote
  {\bibinfo {title} {Nonclassicality of bright
  {Greenberger}-{Horne}-{Zeilinger}--like radiation of an optical parametric
  source},}\ }\href {\doibase 10.1103/PhysRevA.103.042226} {\bibfield
  {journal} {\bibinfo  {journal} {Phys. Rev. A}\ }\textbf {\bibinfo {volume}
  {103}},\ \bibinfo {pages} {042226} (\bibinfo {year}
  {2021}{\natexlab{b}})}\BibitemShut {NoStop}%
\bibitem [{\citenamefont {Chang}\ \emph {et~al.}(2020)\citenamefont {Chang},
  \citenamefont {Sab\'{\i}n}, \citenamefont {Forn-D\'{\i}az}, \citenamefont
  {Quijandr\'{\i}a}, \citenamefont {Vadiraj}, \citenamefont {Nsanzineza},
  \citenamefont {Johansson},\ and\ \citenamefont {Wilson}}]{3phot_exp}%
  \BibitemOpen
  \bibfield  {author} {\bibinfo {author} {\bibfnamefont {C.~W.~Sandbo}\
  \bibnamefont {Chang}}, \bibinfo {author} {\bibfnamefont {Carlos}\
  \bibnamefont {Sab\'{\i}n}}, \bibinfo {author} {\bibfnamefont
  {P.}~\bibnamefont {Forn-D\'{\i}az}}, \bibinfo {author} {\bibfnamefont
  {Fernando}\ \bibnamefont {Quijandr\'{\i}a}}, \bibinfo {author} {\bibfnamefont
  {A.~M.}\ \bibnamefont {Vadiraj}}, \bibinfo {author} {\bibfnamefont
  {I.}~\bibnamefont {Nsanzineza}}, \bibinfo {author} {\bibfnamefont
  {G.}~\bibnamefont {Johansson}}, \ and\ \bibinfo {author} {\bibfnamefont
  {C.~M.}\ \bibnamefont {Wilson}},\ }\bibfield  {title} {\enquote {\bibinfo
  {title} {Observation of three-photon spontaneous parametric down-conversion
  in a superconducting parametric cavity},}\ }\href {\doibase
  10.1103/PhysRevX.10.011011} {\bibfield  {journal} {\bibinfo  {journal} {Phys.
  Rev. X}\ }\textbf {\bibinfo {volume} {10}},\ \bibinfo {pages} {011011}
  (\bibinfo {year} {2020})}\BibitemShut {NoStop}%
\bibitem [{\citenamefont {T\'oth}\ and\ \citenamefont
  {G\"uhne}(2005)}]{GHZ:Indicator}%
  \BibitemOpen
  \bibfield  {author} {\bibinfo {author} {\bibfnamefont {G\'eza}\ \bibnamefont
  {T\'oth}}\ and\ \bibinfo {author} {\bibfnamefont {Otfried}\ \bibnamefont
  {G\"uhne}},\ }\bibfield  {title} {\enquote {\bibinfo {title} {Detecting
  genuine multipartite entanglement with two local measurements},}\ }\href
  {\doibase 10.1103/PhysRevLett.94.060501} {\bibfield  {journal} {\bibinfo
  {journal} {Phys. Rev. Lett.}\ }\textbf {\bibinfo {volume} {94}},\ \bibinfo
  {pages} {060501} (\bibinfo {year} {2005})}\BibitemShut {NoStop}%
\bibitem [{\citenamefont {Yu}\ \emph {et~al.}(2003)\citenamefont {Yu},
  \citenamefont {Pan}, \citenamefont {Chen},\ and\ \citenamefont
  {Zhang}}]{Necessary}%
  \BibitemOpen
  \bibfield  {author} {\bibinfo {author} {\bibfnamefont {Sixia}\ \bibnamefont
  {Yu}}, \bibinfo {author} {\bibfnamefont {Jian-Wei}\ \bibnamefont {Pan}},
  \bibinfo {author} {\bibfnamefont {Zeng-Bing}\ \bibnamefont {Chen}}, \ and\
  \bibinfo {author} {\bibfnamefont {Yong-De}\ \bibnamefont {Zhang}},\
  }\bibfield  {title} {\enquote {\bibinfo {title} {Comprehensive test of
  entanglement for two-level systems via the indeterminacy relationship},}\
  }\href {\doibase 10.1103/PhysRevLett.91.217903} {\bibfield  {journal}
  {\bibinfo  {journal} {Phys. Rev. Lett.}\ }\textbf {\bibinfo {volume} {91}},\
  \bibinfo {pages} {217903} (\bibinfo {year} {2003})}\BibitemShut {NoStop}%
\bibitem [{\citenamefont {Mermin}(1990{\natexlab{b}})}]{Mermin:inequality}%
  \BibitemOpen
  \bibfield  {author} {\bibinfo {author} {\bibfnamefont {N.~David}\
  \bibnamefont {Mermin}},\ }\bibfield  {title} {\enquote {\bibinfo {title}
  {Extreme quantum entanglement in a superposition of macroscopically distinct
  states},}\ }\href {\doibase 10.1103/PhysRevLett.65.1838} {\bibfield
  {journal} {\bibinfo  {journal} {Phys. Rev. Lett.}\ }\textbf {\bibinfo
  {volume} {65}},\ \bibinfo {pages} {1838--1840} (\bibinfo {year}
  {1990}{\natexlab{b}})}\BibitemShut {NoStop}%
\bibitem [{\citenamefont {Menssen}\ \emph {et~al.}(2017)\citenamefont
  {Menssen}, \citenamefont {Jones}, \citenamefont {Metcalf}, \citenamefont
  {Tichy}, \citenamefont {Barz}, \citenamefont {Kolthammer},\ and\
  \citenamefont {Walmsley}}]{Walmsley17}%
  \BibitemOpen
  \bibfield  {author} {\bibinfo {author} {\bibfnamefont {Adrian~J.}\
  \bibnamefont {Menssen}}, \bibinfo {author} {\bibfnamefont {Alex~E.}\
  \bibnamefont {Jones}}, \bibinfo {author} {\bibfnamefont {Benjamin~J.}\
  \bibnamefont {Metcalf}}, \bibinfo {author} {\bibfnamefont {Malte~C.}\
  \bibnamefont {Tichy}}, \bibinfo {author} {\bibfnamefont {Stefanie}\
  \bibnamefont {Barz}}, \bibinfo {author} {\bibfnamefont {W.~Steven}\
  \bibnamefont {Kolthammer}}, \ and\ \bibinfo {author} {\bibfnamefont {Ian~A.}\
  \bibnamefont {Walmsley}},\ }\bibfield  {title} {\enquote {\bibinfo {title}
  {Distinguishability and many-particle interference},}\ }\href {\doibase
  10.1103/PhysRevLett.118.153603} {\bibfield  {journal} {\bibinfo  {journal}
  {Phys. Rev. Lett.}\ }\textbf {\bibinfo {volume} {118}},\ \bibinfo {pages}
  {153603} (\bibinfo {year} {2017})}\BibitemShut {NoStop}%
\bibitem [{\citenamefont {Thekkadath}\ \emph {et~al.}(2020)\citenamefont
  {Thekkadath}, \citenamefont {Phillips}, \citenamefont {Bulmer}, \citenamefont
  {Clements}, \citenamefont {Eckstein}, \citenamefont {Bell}, \citenamefont
  {Lugani}, \citenamefont {Wolterink}, \citenamefont {Lita}, \citenamefont
  {Nam}, \citenamefont {Gerrits}, \citenamefont {Wade},\ and\ \citenamefont
  {Walmsley}}]{Walmsley20}%
  \BibitemOpen
  \bibfield  {author} {\bibinfo {author} {\bibfnamefont {G.~S.}\ \bibnamefont
  {Thekkadath}}, \bibinfo {author} {\bibfnamefont {D.~S.}\ \bibnamefont
  {Phillips}}, \bibinfo {author} {\bibfnamefont {J.~F.~F.}\ \bibnamefont
  {Bulmer}}, \bibinfo {author} {\bibfnamefont {W.~R.}\ \bibnamefont
  {Clements}}, \bibinfo {author} {\bibfnamefont {A.}~\bibnamefont {Eckstein}},
  \bibinfo {author} {\bibfnamefont {B.~A.}\ \bibnamefont {Bell}}, \bibinfo
  {author} {\bibfnamefont {J.}~\bibnamefont {Lugani}}, \bibinfo {author}
  {\bibfnamefont {T.~A.~W.}\ \bibnamefont {Wolterink}}, \bibinfo {author}
  {\bibfnamefont {A.}~\bibnamefont {Lita}}, \bibinfo {author} {\bibfnamefont
  {S.~W.}\ \bibnamefont {Nam}}, \bibinfo {author} {\bibfnamefont
  {T.}~\bibnamefont {Gerrits}}, \bibinfo {author} {\bibfnamefont {C.~G.}\
  \bibnamefont {Wade}}, \ and\ \bibinfo {author} {\bibfnamefont {I.~A.}\
  \bibnamefont {Walmsley}},\ }\bibfield  {title} {\enquote {\bibinfo {title}
  {Tuning between photon-number and quadrature measurements with weak-field
  homodyne detection},}\ }\href {\doibase 10.1103/PhysRevA.101.031801}
  {\bibfield  {journal} {\bibinfo  {journal} {Phys. Rev. A}\ }\textbf {\bibinfo
  {volume} {101}},\ \bibinfo {pages} {031801} (\bibinfo {year}
  {2020})}\BibitemShut {NoStop}%
\bibitem [{\citenamefont {He}\ \emph {et~al.}(2011)\citenamefont {He},
  \citenamefont {Reid}, \citenamefont {Vaughan}, \citenamefont {Gross},
  \citenamefont {Oberthaler},\ and\ \citenamefont
  {Drummond}}]{Stokes:original}%
  \BibitemOpen
  \bibfield  {author} {\bibinfo {author} {\bibfnamefont {Q.~Y.}\ \bibnamefont
  {He}}, \bibinfo {author} {\bibfnamefont {M.~D.}\ \bibnamefont {Reid}},
  \bibinfo {author} {\bibfnamefont {T.~G.}\ \bibnamefont {Vaughan}}, \bibinfo
  {author} {\bibfnamefont {C.}~\bibnamefont {Gross}}, \bibinfo {author}
  {\bibfnamefont {M.}~\bibnamefont {Oberthaler}}, \ and\ \bibinfo {author}
  {\bibfnamefont {P.~D.}\ \bibnamefont {Drummond}},\ }\bibfield  {title}
  {\enquote {\bibinfo {title} {{Einstein}-{Podolsky}-{Rosen} entanglement
  strategies in two-well {Bose}-{Einstein} condensates},}\ }\href {\doibase
  10.1103/PhysRevLett.106.120405} {\bibfield  {journal} {\bibinfo  {journal}
  {Phys. Rev. Lett.}\ }\textbf {\bibinfo {volume} {106}},\ \bibinfo {pages}
  {120405} (\bibinfo {year} {2011})}\BibitemShut {NoStop}%
\bibitem [{\citenamefont {Wasak}\ and\ \citenamefont
  {Chwede{\'n}czuk}(2018)}]{NonCL_BEC}%
  \BibitemOpen
  \bibfield  {author} {\bibinfo {author} {\bibfnamefont {Tomasz}\ \bibnamefont
  {Wasak}}\ and\ \bibinfo {author} {\bibfnamefont {Jan}\ \bibnamefont
  {Chwede{\'n}czuk}},\ }\bibfield  {title} {\enquote {\bibinfo {title} {Bell
  inequality, {Einstein}-{Podolsky}-{Rosen} steering, and quantum metrology
  with spinor {Bose}-{Einstein} condensates},}\ }\href@noop {} {\bibfield
  {journal} {\bibinfo  {journal} {Physical review letters}\ }\textbf {\bibinfo
  {volume} {120}},\ \bibinfo {pages} {140406} (\bibinfo {year}
  {2018})}\BibitemShut {NoStop}%
\bibitem [{\citenamefont {Das}\ \emph {et~al.}(2022{\natexlab{a}})\citenamefont
  {Das}, \citenamefont {Karczewski}, \citenamefont {Mandarino}, \citenamefont
  {Markiewicz}, \citenamefont {Woloncewicz},\ and\ \citenamefont
  {{\.{Z}}ukowski}}]{1stPaper}%
  \BibitemOpen
  \bibfield  {author} {\bibinfo {author} {\bibfnamefont {Tamoghna}\
  \bibnamefont {Das}}, \bibinfo {author} {\bibfnamefont {Marcin}\ \bibnamefont
  {Karczewski}}, \bibinfo {author} {\bibfnamefont {Antonio}\ \bibnamefont
  {Mandarino}}, \bibinfo {author} {\bibfnamefont {Marcin}\ \bibnamefont
  {Markiewicz}}, \bibinfo {author} {\bibfnamefont {Bianka}\ \bibnamefont
  {Woloncewicz}}, \ and\ \bibinfo {author} {\bibfnamefont {Marek}\ \bibnamefont
  {{\.{Z}}ukowski}},\ }\bibfield  {title} {\enquote {\bibinfo {title}
  {Wave{\textendash}particle complementarity: detecting violation of local
  realism with photon-number resolving weak-field homodyne measurements},}\
  }\href {\doibase 10.1088/1367-2630/ac54c8} {\bibfield  {journal} {\bibinfo
  {journal} {New Journal of Physics}\ }\textbf {\bibinfo {volume} {24}},\
  \bibinfo {pages} {033017} (\bibinfo {year} {2022}{\natexlab{a}})}\BibitemShut
  {NoStop}%
\bibitem [{\citenamefont {Das}\ \emph {et~al.}(2022{\natexlab{b}})\citenamefont
  {Das}, \citenamefont {Karczewski}, \citenamefont {Mandarino}, \citenamefont
  {Markiewicz}, \citenamefont {Woloncewicz},\ and\ \citenamefont
  {Żukowski}}]{1stPLA}%
  \BibitemOpen
  \bibfield  {author} {\bibinfo {author} {\bibfnamefont {Tamoghna}\
  \bibnamefont {Das}}, \bibinfo {author} {\bibfnamefont {Marcin}\ \bibnamefont
  {Karczewski}}, \bibinfo {author} {\bibfnamefont {Antonio}\ \bibnamefont
  {Mandarino}}, \bibinfo {author} {\bibfnamefont {Marcin}\ \bibnamefont
  {Markiewicz}}, \bibinfo {author} {\bibfnamefont {Bianka}\ \bibnamefont
  {Woloncewicz}}, \ and\ \bibinfo {author} {\bibfnamefont {Marek}\ \bibnamefont
  {Żukowski}},\ }\bibfield  {title} {\enquote {\bibinfo {title} {Remarks about
  {Bell}-nonclassicality of a single photon},}\ }\href {\doibase
  https://doi.org/10.1016/j.physleta.2022.128031} {\bibfield  {journal}
  {\bibinfo  {journal} {Physics Letters A}\ }\textbf {\bibinfo {volume}
  {435}},\ \bibinfo {pages} {128031} (\bibinfo {year}
  {2022}{\natexlab{b}})}\BibitemShut {NoStop}%
\bibitem [{\citenamefont {Das}\ \emph {et~al.}(2022{\natexlab{c}})\citenamefont
  {Das}, \citenamefont {Karczewski}, \citenamefont {Mandarino}, \citenamefont
  {Markiewicz}, \citenamefont {Woloncewicz},\ and\ \citenamefont
  {{\.{Z}}ukowski}}]{CommentDun}%
  \BibitemOpen
  \bibfield  {author} {\bibinfo {author} {\bibfnamefont {Tamoghna}\
  \bibnamefont {Das}}, \bibinfo {author} {\bibfnamefont {Marcin}\ \bibnamefont
  {Karczewski}}, \bibinfo {author} {\bibfnamefont {Antonio}\ \bibnamefont
  {Mandarino}}, \bibinfo {author} {\bibfnamefont {Marcin}\ \bibnamefont
  {Markiewicz}}, \bibinfo {author} {\bibfnamefont {Bianka}\ \bibnamefont
  {Woloncewicz}}, \ and\ \bibinfo {author} {\bibfnamefont {Marek}\ \bibnamefont
  {{\.{Z}}ukowski}},\ }\bibfield  {title} {\enquote {\bibinfo {title} {Comment
  on `single particle nonlocality with completely independent reference
  states'},}\ }\href {\doibase 10.1088/1367-2630/ac55b1} {\bibfield  {journal}
  {\bibinfo  {journal} {New Journal of Physics}\ }\textbf {\bibinfo {volume}
  {24}},\ \bibinfo {pages} {038001} (\bibinfo {year}
  {2022}{\natexlab{c}})}\BibitemShut {NoStop}%
\bibitem [{\citenamefont {Das}\ \emph {et~al.}(2021)\citenamefont {Das},
  \citenamefont {Karczewski}, \citenamefont {Mandarino}, \citenamefont
  {Markiewicz},\ and\ \citenamefont {Żukowski}}]{3rdPaper}%
  \BibitemOpen
  \bibfield  {author} {\bibinfo {author} {\bibfnamefont {Tamoghna}\
  \bibnamefont {Das}}, \bibinfo {author} {\bibfnamefont {Marcin}\ \bibnamefont
  {Karczewski}}, \bibinfo {author} {\bibfnamefont {Antonio}\ \bibnamefont
  {Mandarino}}, \bibinfo {author} {\bibfnamefont {Marcin}\ \bibnamefont
  {Markiewicz}}, \ and\ \bibinfo {author} {\bibfnamefont {Marek}\ \bibnamefont
  {Żukowski}},\ }\bibfield  {title} {\enquote {\bibinfo {title} {Optimal
  interferometry for bell$-$nonclassicality by a vacuum$-$one$-$photon
  qubit},}\ }\href {\doibase 10.48550/ARXIV.2109.10170} {\  (\bibinfo {year}
  {2021}),\ 10.48550/ARXIV.2109.10170}\BibitemShut {NoStop}%
\bibitem [{\citenamefont {\ifmmode~\dot{Z}\else \.{Z}\fi{}ukowski}\ \emph
  {et~al.}(2016)\citenamefont {\ifmmode~\dot{Z}\else \.{Z}\fi{}ukowski},
  \citenamefont {Wie\ifmmode~\acute{s}\else \'{s}\fi{}niak},\ and\
  \citenamefont {Laskowski}}]{Zukowski:Stokes}%
  \BibitemOpen
  \bibfield  {author} {\bibinfo {author} {\bibfnamefont {Marek}\ \bibnamefont
  {\ifmmode~\dot{Z}\else \.{Z}\fi{}ukowski}}, \bibinfo {author} {\bibfnamefont
  {Marcin}\ \bibnamefont {Wie\ifmmode~\acute{s}\else \'{s}\fi{}niak}}, \ and\
  \bibinfo {author} {\bibfnamefont {Wies\l{}aw}\ \bibnamefont {Laskowski}},\
  }\bibfield  {title} {\enquote {\bibinfo {title} {Bell inequalities for
  quantum optical fields},}\ }\href {\doibase 10.1103/PhysRevA.94.020102}
  {\bibfield  {journal} {\bibinfo  {journal} {Phys. Rev. A}\ }\textbf {\bibinfo
  {volume} {94}},\ \bibinfo {pages} {020102} (\bibinfo {year}
  {2016})}\BibitemShut {NoStop}%
\bibitem [{\citenamefont {{\.Z}ukowski}\ \emph {et~al.}(1997)\citenamefont
  {{\.Z}ukowski}, \citenamefont {Zeilinger},\ and\ \citenamefont
  {Horne}}]{multiport}%
  \BibitemOpen
  \bibfield  {author} {\bibinfo {author} {\bibfnamefont {Marek}\ \bibnamefont
  {{\.Z}ukowski}}, \bibinfo {author} {\bibfnamefont {Anton}\ \bibnamefont
  {Zeilinger}}, \ and\ \bibinfo {author} {\bibfnamefont {Michael~A}\
  \bibnamefont {Horne}},\ }\bibfield  {title} {\enquote {\bibinfo {title}
  {Realizable higher-dimensional two-particle entanglements via multiport beam
  splitters},}\ }\href@noop {} {\bibfield  {journal} {\bibinfo  {journal}
  {Physical Review A}\ }\textbf {\bibinfo {volume} {55}},\ \bibinfo {pages}
  {2564} (\bibinfo {year} {1997})}\BibitemShut {NoStop}%
\end{thebibliography}
\end{document}